\documentclass[compsoc,conference,10pt,times]{IEEEtran}

\usepackage{cite}
\usepackage{amsmath,amssymb,amsfonts}

\usepackage{textcomp}
\usepackage{bmpsize}
\usepackage{xcolor}
\usepackage{lipsum}
\usepackage{enumitem}
\usepackage[colorlinks=true,urlcolor=black]{hyperref}
\def\BibTeX{{\rm B\kern-.05em{\sc i\kern-.025em b}\kern-.08em
    T\kern-.1667em\lower.7ex\hbox{E}\kern-.125emX}}
    
\usepackage{tikz}

\usepackage{booktabs}
\usepackage{amsmath}
\usepackage{amsmath,amssymb,amsfonts}
\usepackage{algorithm}
\usepackage{algpseudocode}

\usepackage{amsthm}

\usepackage{comment}

\usepackage{xcolor}
\usepackage{listings}
\lstdefinestyle{tcstyle}{
  backgroundcolor=\color{gray!10},
  basicstyle=\ttfamily\small,
  keywordstyle=\color{blue}\bfseries,
  commentstyle=\color{green!60!black}\itshape,
  stringstyle=\color{orange},
  identifierstyle=\color{black},
  breaklines=true,
  breakatwhitespace=true,
  frame=single,
  framerule=0.5pt,
  rulecolor=\color{gray!50},
  numbers=left,
  numberstyle=\tiny\color{gray},
  numbersep=6pt,
  xleftmargin=5pt,
  xrightmargin=5pt,
  showstringspaces=false,
  columns=fullflexible
}

\usetikzlibrary{tikzmark}
\definecolor{calloutred}{HTML}{FF7461}
\tikzset{mycircled/.style={circle,draw,inner sep=0.1em,line width=0.04em}}
\def\callout#1{\tikzmarknode[mycircled,draw=calloutred,fill=calloutred]{t1}{\textcolor{white}{#1}}}

\def\CAPIO#1{\textsc{CAP}\-\textsc{IO}}

\begin{document}

\title{\CAPIO{}: Safe Kernel-Bypass of Commodity Devices using Capabilities}

\author{
    \IEEEauthorblockN{Friedrich Doku, Jonathan Laughton, Nick Wanninger, and Peter Dinda}
    \IEEEauthorblockA{Northwestern University \\
    \{friedy, ncw\}@u.northwestern.edu, laughtonjonathan@gmail.com, pdinda@northwestern.edu}
}

\ifx\editingmarks\undefined
\newcommand{\myremark}[3]{}
\else
\newcommand{\myremark}[3]{
\refstepcounter{remark}
\[
\left\{
\sf
\parbox{0.8\columnwidth}
{
{\bf {#1}'s remark~\arabic{section}.\arabic{remark}:}
{#3}
}
\right.
\]
\marginpar{\bf {#2}.~\arabic{section}.\arabic{remark}}
}
\fi

\definecolor{johncolor}{HTML}{6FD3BC}
\newcommand{\jremark}[1]{\textcolor{johncolor}{\myremark{John}{J}{#1}}}
\newcommand{\fremark}[1]{\textcolor{blue}{\myremark{Friedy}{F}{#1}}}

\def\jnotepad#1{$\rightarrow$ {\color{johncolor} \bf JL: #1} }
\def\fnotepad#1{$\rightarrow$ {\color{blue} \bf FD: #1} }
\def\ncw#1{$\rightarrow$ {\color{teal} \bf NCW: #1} }
\def\pad#1{$\rightarrow$ {\color{brown} \bf PAD: #1} }

\def\todocite#1{[{\color{red} \bf CITE}]}

\def\point#1{{\bf #1:}}

\def\secref#1{\S\ref{#1}}

\def\smallcolfig#1{\includegraphics[width=2.0in]{#1}}
\def\colfig#1{\includegraphics[width=0.8\linewidth]{#1}}
\def\tricolfig#1{\includegraphics[width=0.33\linewidth]{#1}}
\def\pagefig#1{\includegraphics[width=5in]{#1}}
\def\suspcolfig#1{\includegraphics[width=2.0in]{#1}}
\def\tffig#1{\includegraphics[width=0.7\linewidth]{#1}}
\def\bigtffig#1{\includegraphics[width=0.75\linewidth]{#1}}
\def\roundfig#1{\includegraphics[width=0.75\linewidth]{#1}}
\def\bigroundfig#1{\includegraphics[width=0.77\linewidth]{#1}}

\maketitle

\begin{abstract}
Securing low-latency I/O in commodity systems forces a fundamental trade-off: rely on the kernel's high overhead mediated interface, or bypass it entirely, exposing sensitive hardware resources to userspace and creating new vulnerabilities. This dilemma stems from a hardware granularity mismatch: standard MMUs operate at page boundaries, making it impossible to selectively expose safe device registers without also exposing the sensitive control registers colocated on the same page.
Existing solutions to driver isolation, enforce an isolation model that cannot protect sub-page device resources.

This paper presents \CAPIO{}, the first architecture to leverage hardware capabilities to enforce fine-grained access control on memory-mapped I/O. Unlike prior page-based protections, \CAPIO{} utilizes unforgeable capabilities to create precise, sub-page ``slices'' of device memory. This mechanism enables the kernel to delegate latency-critical hardware access to userspace applications while strictly preventing interaction with co-located privileged registers.

We implement \CAPIO{} based on CHERI on the ARM Morello platform and demonstrate a proof-of-concept safe-access driver for a commodity network card which was not originally designed for kernel bypass.
We demonstrate that \CAPIO{} achieves the latency improvements of kernel bypass while enforcing byte-level access control of privileged resources.
\end{abstract}

\maketitle

\section{Introduction}

The kernel often stands in the way of low-latency access to device hardware through expensive ring transitions, abstractions, and memory copy overheads~\cite{bypassd, ipc_stuart, exokernel-engler}.
Modern commodity systems like Linux or FreeBSD expose their devices through a system call interface, requiring applications to cross the user/kernel boundary for nearly every I/O operation.
Because the kernel is ultimately the only entity interacting with the device, this {\em kernel-mediated device access} is simple, abstract, general, and {\em safe.}  A userspace process cannot use the device to mount an attack on the kernel or another process.    However, kernel-mediated device access imposes a significant performance overhead.  This overhead is particularly problematic for high-performance I/O workloads where microseconds matter, such as in high-frequency trading or high-performance computing. 

In contrast, {\em kernel-bypass device access} allows the userspace process to directly drive the device, choosing performance over safety.    High-end network cards such as NVIDIA Infiniband or AMD Solarflare support user-level access to give developer direct control over the entire network stack and the card itself to create a bespoke design that squeezes out every last cycle.
Kernel-bypass device access is desirable beyond networking. For example, databases aiming for maximum performance directly interact with storage controllers according to their needs \cite{kernel-bypass-is-not-optional}.

Ideally, we would not have to choose between safety and performance.   A core problem of achieving {\em safe kernel-bypass device access} is that of exposing device state and registers to the userspace in a manner that cannot violate safety and security requirements.   
With existing systems, userspace access to the device is achieved by mapping the device's resources into userspace virtual addresses with page granularity (typically 4KB).
As a result, registers and memory that happen to co-reside in a physical page frame in the hardware design are either {\em all} fully readable and writable or {\em none} of them are.   As an example of this {\em protection granularity problem}, a status register that userspace might legitimately and safely read might co-exist in the same page frame with a register that a malicious userspace could use to enable unfettered access to all of physical memory through DMA transfers.

High-end network cards achieve safe kernel-bypass device access by essentially providing an additional  {\em alternative user-safe interface} that can instead be mapped into the userspace, with the full-functionality interface being reserved for the kernel.  A common model is to leverage hardware features like SR-IOV~\cite{high-perf-sriov} along with proprietary hardware mechanisms to expose many virtual device queues which can be safely modified and updated by the userspace applications.   
For example, on AMD Solarflare cards, the user creates a ``virtual interface'' which allows the application to send and receive raw Ethernet frames over various ports through a virtual command queue safely mapped into their address space correctly \cite{onload}.
The hardware then pools these queue requests and dispatches them as needed from all virtual interfaces.

The alternative user-safe interface approach is extremely complicated to implement on a device like a network card. It is also overkill in the common case where only a single userspace process needs direct access to the device. The reality is that commodity devices, including network cards, storage controllers, GPUs, accelerators and others, do not provide an alternative user-safe interface, and thus cannot be safely exposed to userspace due to the protection granularity problem.  

Device-independent hardware support for protection, such as trusted execution environments (TEEs) and virtualization do not solve the protection granularity problem.    TEEs, such as Intel SGX and ARM TrustZone, provide isolation for code and data, but do so at a page granularity, generally enforced by the system MMU and IOMMU. Porting drivers or applications to a TEE OS is also a major engineering effort. Virtualization also has the same issue because, once again, protections for memory-mapped device resources operate at the page granularity.   Indeed, this is such an issue that some core-local devices like the x64 APIC provide model-specific register (MSR)-level access to allow per-MSR access to be extended to the guest, for example to allow the guest to EOI an interrupt, but not launch one.   Only a very tiny set of devices are core-local; per-MSR access control is irrelavent to the vast majority of devices.  Of course, a trap-and-emulate approach also allows finer control, but each trap is an entry into the kernel, obviating the point of kernel-bypass device access in the first place.

In this work, we identify the protection granularity problem as the key challenge to achieving safe kernel-bypass device access, and we demonstrate that the problem can be readily solved on a system that provides hardware capabilities.    More specifically, we show how to use a CHERI-enabled~\cite{CHERI} memory system to solve the problem.   CHERI (Capability Hardware Enhanced RISC Instructions) is an architectural extension that replaces traditional integer-based pointers with unforgeable, hardware-enforced capabilities.
These capabilities are bounded, tagged references that enforce spatial safety by defining precise access permissions for memory regions, including the base address, length, and allowed operations (read, write, execute).

Capabilities offer fine-grained, {\em sub-page} (down to byte-granularity) access control enforced by the processor with almost no changes required at the application level~\cite{CHERI}.
With CHERI, the kernel simply maps the entirety of a device's resources into userspace virtual addresses without fear---absent the correct capabilities the application cannot dereference these addresses in any way.
The kernel doles out capabilities for specific {\em byte slices} of the device as it sees fit, tightly controlling the access to the device's resources  in order to enforce a security policy.
This lets the kernel retain exclusive control over privileged operations like DMA configuration and interrupt management at the ``control plane'', but defer to the userspace application for ``data plane'' activities like sending and receiving packets according to its bespoke design. 
In this model, capabilities serve as unforgeable I/O access tokens: applications cannot forge or expand capabilities to access restricted device registers, even if they exploit memory safety vulnerabilities.

Our approach, \CAPIO{}, consists of three key components: the kernel stub, the device manifest, and the slicer.  
The kernel stub is responsible for mapping the device resources into the userspace and for handling ``control plane'' operations that the userspace cannot be trusted to do.  The device manifest is essentially an access policy for the device resources, indicating how each range of MMIO addresses mapped by the kernel stub into the userspace is allowed to be accessed.   Finally, the slicer transforms the manifest into bounded capabilities that are returned to the user process and can be used by it to access specific runs of device registers and other resources.

Critically, \CAPIO{}, works with \textit{any} existing, unmodifed commodity device that is accessed via MMIO.   \CAPIO{} also has the potential to simplify the design of future high-performance devices, eliminating the need for them to have an alternative user-safe interface.   By addressing the protection granularity problem through hardware capabilities implemented in the processor, \CAPIO{} decouples safety from specific device features. It also transforms the kernel from being a data-path bottleneck into being a precise policy administrator, enabling a new class of performant userspace drivers that are safe to use.

We present the design, implementation, and evaluation of a \CAPIO{} prototype for the ARM Morello platform that provides an existence proof that safe kernel-bypass device access to commodity devices can not only be achieved, but can be performant.
Our target device is an off-the-shelf Intel 82754L (e1000e) network card, which was \textit{never designed for kernel bypass}.  We are able to provide safe kernel-bypass device access to this card for a FreeBSD process that includes a simple user-level embedded network stack (LwIP) that has a driver for it.  This \CAPIO{} setup shows considerable reductions in packet round-trip latency compared to an alternative that uses safe kernel-mediated decive access via the FreeBSD network stack.   These microbenchmark results are reinforced by application-level latency reductions of about 30\% for Memcached driven by a YCSB workload. These results suggest that any overheads in the \CAPIO{} approach to safe kernel-bypass device access are likely to be dominated by the performance gains of making user-level device access possible.

{\bf Our specific contributions are:}
\begin{enumerate}

\item We identify the protection granularity problem as key to achieving safe kernel-bypass device access for unmodified commodity devices.
\item We show how to address the protection granularity problem through the mechanism of a hardware capability system.
\item We describe \CAPIO{}, the first architecture to leverage hardware capabilities for enforcing I/O access control at the granularity of individual device registers instead of pages of registers.   In our design, kernel/user cooperation results in exposing exactly the safe subset of device resources required to achieve an application goal.  We use low-latency networking as an example goal (\secref{sec:design}).
\item We give a detailed breakdown of the design and implementation of the \CAPIO{} prototype for the ARM Morello platform, and exemplify the design using the e1000e network card (\secref{sec:implementation}).
\item We create a user-level driver for the e1000e and couple it with a simple user-level network stack to create a complete example of safe kernel-bypass user-level networking based on the \CAPIO{} approach.
\item We evaluate the resulting user-level networking setup, showing that it does indeed reduce microbenchmarked UDP latency by up to 46\% and application TCP latency by up to 19\% (\secref{sec:evaluation}).
\end{enumerate}
We believe that the \CAPIO{} approach is quite general and shows an unexpected benefit of the adoption of hardware capability machines.  We also think that our \CAPIO{} prototype generalizes beyond the e1000e or even network cards.   Most commodity devices of note operate on a similar descriptor-and-queue model~\cite{g-nvme2019spec, g-spdk2024nvme, g-vfio2023}, and, as a result, even the specific scheme we used for the e1000e is likely to be applicable across a large part of the hardware landscape.  Our code will be made available on publication of this paper.

\section{Background and Related Work}
\label{sec:related}

In the context of this work, we define \textit{Trusted I/O} as the property by which input and output exchanges between an application and a hardware device are protected such that only authorized parties may initiate or observe transactions, and data cannot be tampered with or intercepted by unprivileged components~\cite{tio-amd, tio-guo, tio-liang, tio-sgx}. Achieving Trusted I/O in commodity systems remains an open challenge, as existing mechanisms generally force a trade-off between access control granularity and raw performance.

Prior architectural research has largely bifurcated into two distinct approaches. \textit{Isolation-centric systems} (e.g., TEEs) prioritize security by enforcing strict logical separation, often at the cost of performance. By leveraging TEEs, we can ensure that the software controlling the device is trustworthy, reducing the need for strict enforcement of access protections on device resources.

Conversely, \textit{Page-based protection mechanisms} (e.g, Virtualization, and Kernel-Bypass) prioritize performance but rely on coarse-grained memory management units. \CAPIO{} occupies a novel intersection, utilizing CHERI capabilities to enforce strict, sub-page access control without the overhead of the former or the insecurity of the latter.

\subsection{Trusted Execution Environments (TEEs)}
Hardware-based TEEs create Trusted I/O paths by isolating driver execution to a protected memory region. Attestation can be used to verify that the device logic accesses registers as expected, but this comes with porting challenges and performance costs.

Intel SGX enclaves are designed primarily for computation and explicitly lack direct access to device MMIO~\cite{tio-liang, tio-sgx}. Solutions like Sgxio~\cite{tio-sgx} and Aurora~\cite{tio-liang} must construct synthetic trusted paths using software proxies. Aurora, for instance, leverages System Management Mode (SMM) to intercept I/O interrupts. While effective for isolation, this introduces significant latency due to the heavyweight nature of SMI handlers and the inability of the OS to preempt SMM execution~\cite{tio-liang}.

ARM TrustZone partitions the processor into Secure and Normal worlds. ``Minimum Viable Device Drivers'' (MVD)~\cite{tio-guo} leverages this to run security-critical driver logic in the Secure World. While MVD reduces engineering effort via record-replay, it still necessitates a synchronous \textit{World Switch}, a hardware context switch involving pipeline flushes, whenever the Normal World interacts with the device~\cite{tio-guo}; this is typically more expensive than a switch from user mode to kernel mode. 

\subsection{Virtualization and IOMMUs}
Hardware virtualization protects system memory from malicious devices by enforcing access control at the bus level.

Systems like SUD (Safe User-level Drivers)~\cite{tio-boyd} run unmodified drivers in userspace, relying on the IOMMU to restrict the device's DMA capabilities to specific memory regions. This successfully isolates the kernel from driver crashes, but relies entirely on page-based protection mechanisms to restrict the driver's view of the device~\cite{tio-boyd}.

AMD SEV-TIO (Trusted I/O) extends memory encryption to the I/O subsystem~\cite{tio-amd}. SEV-TIO allows guests to share encrypted memory directly with trusted devices, eliminating software bounce buffers. It leverages the TDISP and IDE standards to enforce isolation at the PCIe link level. However, SEV-TIO targets the virtualization model, assigning devices (or Virtual Functions) to entire Virtual Machines rather than individual applications or processes.

\subsection{Kernel-Bypass Architectures}
To eliminate system call overhead, high-performance applications utilize kernel-bypass frameworks such as DPDK~\cite{dpdk}, SPDK~\cite{spdk}, and netmap~\cite{netmap}. These architectures map device MMIO regions and DMA buffers directly into the application's virtual address space. While effective for latency, these approaches historically necessitate a binary trade-off: to gain direct access, the application must be fully trusted (often running as root) because the hardware provides no mechanism to restrict access to specific registers within the mapped region.

\subsection{The Protection Granularity Problem}
\label{granularity}
While the approaches above differ in implementation, SUD, SEV-TIO, and DPDK all share a fundamental limitation: they rely on the IOMMU or MMU for protection, which operates at page granularity (typically 4KB).

Commodity device register maps are rarely designed with page-aligned security boundaries. Privileged configuration registers (e.g., interrupt masking, DMA root pointers) frequently reside on the same physical page as the safe I/O registers (e.g., packet tail pointers) required for operation~\cite{tio-boyd}. Consequently, page-based systems cannot expose performance-critical I/O paths to userspace without inadvertently exposing the control registers, allowing a malicious application to reconfigure the device or trigger interrupts.

\CAPIO{} resolves this tension by overlaying CHERI capabilities on top of these memory-mapped regions. By bringing hardware-enforced isolation down to the pointer level, \CAPIO{} can safely expose specific sub-page register ranges to userspace, achieving the performance of kernel-bypass and the isolation of a TEE without the granularity limitations of legacy hardware.

\subsection{CHERI}

CHERI extends conventional processor architectures with hardware-enforced capabilities, unforgeable tokens that grant specific access rights to bounded memory regions~\cite{CHERI}. Unlike traditional pointers, which are simply integer addresses that can be arbitrarily forged, capabilities are protected by hardware and encode several critical attributes: a base address and length defining accessible memory bounds, permissions specifying allowed operations (read, write, execute), a tag bit indicating validity, and optional object type metadata.

The processor enforces these attributes on every memory access. Attempts to access memory outside a capability's bounds, perform unpermitted operations, or use an untagged value as a capability trigger hardware exceptions. This provides \textit{spatial memory safety} by preventing buffer overflows, out-of-bounds accesses, and pointer arithmetic errors. Bounds cannot be expanded, any attempt to create a capability exceeding its parent's range fails or produces an untagged value.

CHERI provides \textit{sealed capabilities}, which are crtical for \CAPIO{}. CHERI's \textit{sealed capabilities} provide unforgeable authentication tokens. Sealing binds a capability to a specific object type and prevents the sealed capability from being dereferenced or modified. Only software possessing the corresponding otype capability can unseal it, creating a protected token proving the holder was authorized by the original sealer.

This enables provenance verification: when an application presents a sealed capability to the kernel, the kernel verifies authenticity by unsealing it with the appropriate otype capability. Success proves the capability was legitimately issued and has not been forged. Even if an attacker exploits kernel vulnerabilities, they cannot forge sealed capabilities without access to the secret otype capability. The sealing mechanism is enforced entirely in hardware through the processor's capability manipulation instructions.

\subsection{The ARM Morello Platform}

ARM Morello is a prototype SoC implementing CHERI capabilities on ARMv8.2-A architecture~\cite{morello}. The platform extends ARM with capability-specific instructions while maintaining backwards compatibility through hybrid mode. Pointer are 128 bits\footnote{With one extra `valid' bit placed in inaccessible memory}, encoding address, compressed bounds, permissions, and a validity tag stored in separate tagged memory.

The Morello board features a quad-core ARM Neoverse N1 processor (up to 2.5 GHz), 64GB DDR4 memory, and standard I/O interfaces including PCIe, USB, and networking. This allows running full-featured operating systems like CheriBSD~\cite{cheribsd} and Morello Linux with actual device drivers. Critically for \CAPIO{}, Morello implements full hardware-enforced sealing with \texttt{seal} and \texttt{unseal} instructions, making sealed capabilities unforgeable even in the presence of kernel vulnerabilities.

\subsection{Software Compartmentalization on CHERI}
Recent work has leveraged CHERI to compartmentalize userspace network stacks. Ferraro et al. ported DPDK and F-Stack to Arm Morello, isolating the TCP/IP stack from the application logic using capability-based VMs (cVMs) \cite{cheri-ferraro}. While this prevents software vulnerabilities (e.g., buffer overflows) within the stack from corrupting the application, it relies on standard DPDK device mapping. Consequently, the isolated network compartment still retains full, coarse-grained access to the device's entire MMIO region, leaving the device vulnerable to malicious reconfiguration if the network compartment is compromised. \CAPIO{} complements this work by enforcing fine-grained access control at the register level, ensuring that even a compromised driver compartment cannot access privileged device control registers.

\section{Design of \CAPIO{}} 
\label{sec:design}
\CAPIO{} addresses the protection granularity problem (\S\ref{granularity}) by fundamentally shifting where access control is enforced. Modern devices often pack unrelated control plane registers and data plane registers on the same page as control plane registers. For example, a device may have the registers that control global interrupt masks on the same page as DMA queues.    

To safely isolate a function, the system must be able to grant access to the safe data plane registers while denying access to the adjacent control plane registers \cite{nickgordon}. 

Traditional approaches rely on the Memory Management Unit (MMU) as the sole enforcement mechanism, conflating address translating with access protection.
As the MMU operates at fixed page granularities (typically) 4 KB, it lacks the precision required to isolate these co-located resources. This ``all-or-nothing'' model forces a serious trade-off: system designers must either sacrifice security by exposing sensitive control registers alongside safe data interfaces, or sacrifice performance by leaving access to only the kernel. To solve this problem we must enforce access control at the granularity of individual registers.

\subsection{Threat Model} 
\label{sec:threat-model}

We consider an environment where the operating system kernel is trusted but userspace applications, including those with direct device access, are potentially malicious or compromised. Our primary security objective is to enforce \textit{Trusted I/O} by strictly confining the driver's access to a subset of device registers defined by the kernel. We aim to ensure that a compromised application cannot leverage coarse-grained memory mappings to access unauthorized control registers, subvert the hardware, or interfere with other devices.

\subsubsection{Attacker Capabilities}
We assume an attacker has full control over a userspace application (the \textit{compromised application}). The attacker can execute arbitrary code with the application's privileges, read and write any memory within its address space.

Crucially, we assume the attacker has been granted \textit{legitimate} access to a specific device resource (e.g., a specific transmit queue). The attacker attempts to abuse this access by exploiting granularity mismatches to access privileged control registers, such as DMA configuration or interrupt masks, that reside on the same physical page as legitimate data plane registers. Furthermore, the attacker may attempt to forge or manipulate capabilities to access unauthorized device regions, or launch confused deputy attacks by passing invalid pointers to the kernel to trigger unintended behavior. 

\subsubsection{Trust Assumptions}
At the hardware layer, we trust the processor to correctly enforce CHERI capability semantics, including bounds checking, permission validation, and the immutability of sealed capabilities. 

At the software layer, we trust the operating system kernel to correctly implement the \CAPIO{} allocation logic and manage the root capabilities for device memory. We assume the kernel itself is compiled with CHERI pure-capability protections, significantly mitigating the risk of memory-safety vulnerabilities within the trusted computing base (TCB).

While the kernel is trusted to define access \textit{policy} manage capabilities, we do \textit{not} trust the kernel to efficiently police every register access in software; that enforcement is offloaded to the hardware capabilities, ensuring minimal overhead checks on the critical path.

\CAPIO{} aims to enforce the following properties:

\begin{enumerate}
    \item \textbf{Sub-Page Device Isolation:} The system must enforce access control at the granularity of individual registers or fields, independent of page boundaries. An application granted access to a data queue must be architecturally prevented from accessing control registers on the same page.
    \label{goal1}
    \item \textbf{Unforgeable Authorization:} Device access rights must be encoded in unforgeable tokens (sealed capabilities). An attacker cannot fabricate these tokens or derive valid capabilities for unauthorized offsets, even with arbitrary code execution in their own address space.
    \label{goal2}

    \item \textbf{I/O Provenance:} The kernel provides a verifiable association between I/O mapping requests and the userspace processes that issue them, preventing unauthorized processes from impersonating valid ones.
    \label{goal3}
    
    \item \textbf{Containment of Userspace Drivers:} A userspace driver with direct hardware access must be strictly confined to its assigned resources. It cannot reset the device, reconfigure global DMA settings, or interfere with other slices of the device assigned to other contexts.
    \label{goal4}

\end{enumerate}

\subsubsection{Out-of-Scope Threats}
We do not consider attacks requiring physical access to the hardware (e.g., bus snooping, cold boot attacks) or side-channel attacks (e.g., Spectre/Meltdown). We assume the physical device hardware behaves according to its specification and is not malicious. We do not address denial-of-service attacks where a driver refuses to process data.

\subsection{Overview of CAPIO}
\begin{figure}[]
\centering
\includegraphics[width=1.0\columnwidth]{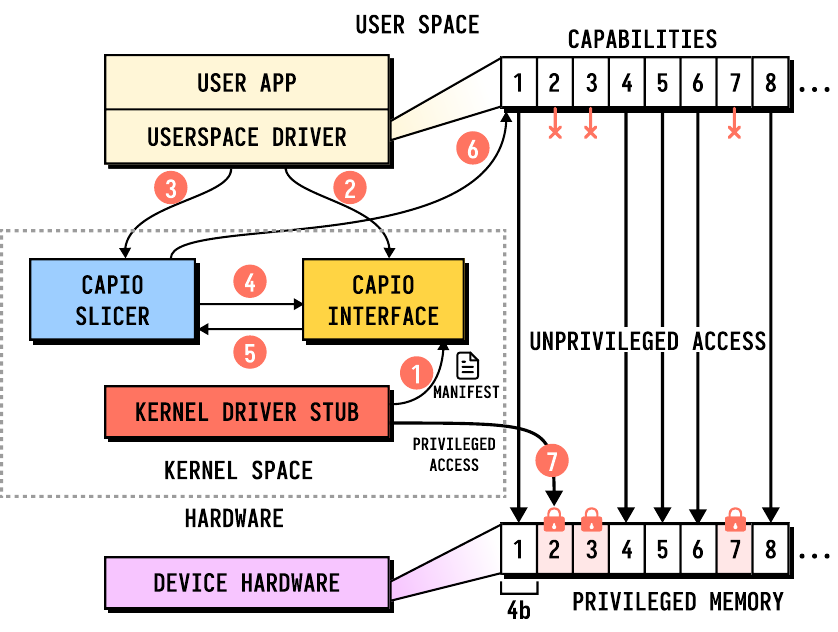}
\caption{\CAPIO{} system architecture. The kernel generates bounded capabilities for the authorized driver, enabling direct access to device registers. A compromised or malicious application attempting to access privileged control registers on the same page is architecturally blocked by CHERI hardware bounds checks, preventing unauthorized device reconfiguration.}
\label{fig:capio_arch}
\end{figure}

In the \CAPIO{} architecture, the MMU handles address translation and process isolation, while CHERI hardware capabilities assume the role of fine-grained access enforcement. 
This separation of concerns allows the kernel to act as a precise policy administrator, defining access rights at the \textbf{byte} level.
Figure~\ref{fig:capio_arch} gives an overview of \CAPIO{} system architecture.

The \CAPIO{} architecture consists of five primary components: a pure capability kernel, the kernel stub, the Interface, the userspace driver, and the Slicer. 
\CAPIO{} requires a \textbf{pure capability kernel}, which serves as the root of trust and governs access to all device resources. 
In \CAPIO{}, the kernel itself runs in pure-capability mode. 
Crucially, it does not allow unprivileged userspace processes to map arbitrary physical memory (e.g, via \verb./dev/mem.).

The life cycle begins when the associated \textbf{kernel stub} is loaded. 
The stub is in charge of allocating the device memory, initializing the device, and defining the \textit{Device Manifest}, a specification of the device's register layout and access permissions to these registers. 
The kernel stub initializes the device, sets up the device memory, and creates shared memory region definitions. 
It then registers these with the \textbf{Interface} \callout{1}. The Interface provides an API to mediate userspace access to devices, and manage mappings defined by the driver.

A \textbf{userspace driver} then attaches to the device through \CAPIO{} \callout{2}. 
The userspace driver is an unprivileged, capability-aware process responsible for the high-level device logic. 
The userspace driver requests mappings to device memory (registers, descriptors, buffers, etc...) through the \textbf{Slicer} \callout{3}. 
The Slicer creates a trust boundary by interposing on the standard \verb.mmap. system call for devices. 
Instead of allowing the userspace driver to map raw physical pages, the Slicer intercepts \verb.mmap. requests, and assigns fine-grained permissions based on the Device Manifest.

To fulfill these requests the Slicer calls into the Interface \callout{4} to request the mapping definitions for a specific device \callout{5}. 
The Slicer provides a runtime policy specifying the exact offsets, lengths, and permissions for each device’s register map. 
It carves a page of registers into slices, each of which is associated with a capability.    The Slicer then creates bounded capabilities based on this information and returns them to userspace \callout{6} for use in the userspace driver.  
Because these capabilities are unforgeable, the driver is architecturally prevented from accessing any system state, such as unrelated registers, that lies outside its explicitly granted slices.

Once this setup is complete, the userspace driver can perform unprivileged accesses through these bounded capabilities with no further kernel interaction in the common (allowed) case.  Such accesses may include device register reads/writes, buffer reads/writes, DMA descriptor reads/writes, etc. 
If the userspace driver needs privileged access (to something for which it has no capabilities), it invokes the kernel stub to perform this access on its behalf, mediated by the kernel in the usual manner \callout{7}.
An example of such privileged access is configuring DMA registers, as allowing the userspace to write them gives unbounded access to all of physical memory.

\section{Implementation of \CAPIO{}} 
\label{sec:implementation}

We implemented \CAPIO{} on the ARM Morello Platform, a functional CHERI extended ARMv8-A SoC \cite{morello}, running CheriBSD \cite{cheribsd}. 
Our implementation consisted of the \textit{Slicer} kernel module, the \CAPIO{} interface, a kernel stub for a 1GbE e1000e network card, a userspace driver for that same card, and the lwip network stack.
This card was chosen as it has a relatively simple device driver model, while maintaining a ``modern''\footnote{Devices programmed through MMIO-based command queues}. interface of queues and descriptors which you use to issue commands like send/receive.
It was also chosen explicitly \textit{because} it is not amenable to safe kernel bypass as privileged registers are placed interleaved with unprivileged ones, and it requires the kernel to manage privileged DMA memory configurations.
We believe that, despite the card's simplicity, its programming interface is in-line with a majority of hardware network interfaces, and even beyond that to devices like disks through AHCI or NVMe\footnote{These interfaces might have more complexity with restricting access to the DMA `paddr' registers, though.}.

\subsection{CheriBSD}
To build \CAPIO{}, we leveraged the CheriBSD operating system, an adaptation of FreeBSD for the Morello architecture \cite{cheribsd}. 
Our prototype requires a kernel capable of managing tagged capability memory. 
CheriBSD provides the necessary \textit{capability-aware} primitives. 
To implement \CAPIO{} the CheriBSD kernel required minimal changes (around $\sim$50 lines of code). 
These were primarily just modifications to the main \verb.mmap. function in the kernel to allow for requests with an extra pointer, and defining a new device \verb.mmap. callback to accept this extra pointer, and providing an entry function for the Slicer. 
Importantly, unrelated \verb.mmap. requests were not broken and compatibility is maintained.

\subsection{Slicer}
The core enforcement mechanism, \textit{Slicer}, was implemented as a CheriBSD kernel module in $\sim$416 lines of C code.
The primary function of the \textit{Slicer} is to allow for the passing of extra information into mmap requests, handle callback registration for a specific character device, and do the slicing of a memory region based on the manifest returned from the callbacks. 
Slicing is done simply by calling Cheri intrinsics to set the bounds and permissions of a given root capability (in this case the whole device memory region) and returning an array of those sliced and permission restricted capabilities to userspace. 
A sealed version of the root capability is returned in the address field of the mmap request, so that memory unmapping is still possible without breaking the security model by returning the original capability.

\subsection{Interface}
The \CAPIO{} interface exists to abstract away the complexities of device access and managing the mapping of device memory. 
Device access is handled through sealed capabilities to create unforgeable tokens that identify the calling process. 
These sealed capabilities are generated on attach and created by the \CAPIO{} interface. 
All future requests (mapping or ioctl calls) require this token to be provided.

\begin{figure}[t]
\centering
\includegraphics[width=1.0\columnwidth]{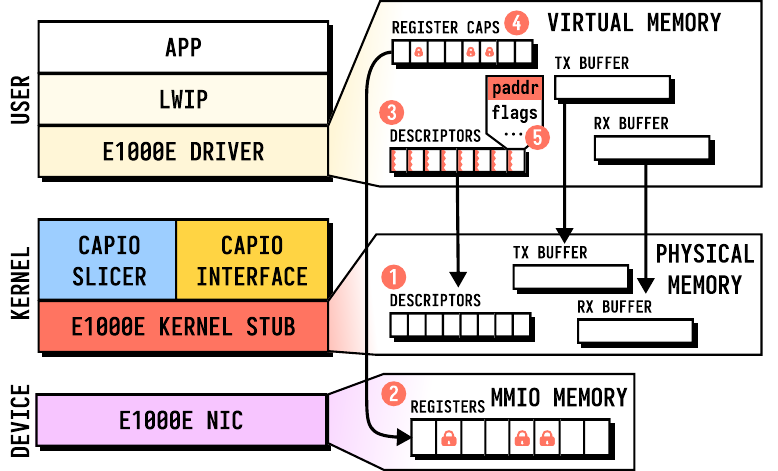}
\caption{\CAPIO{} implementation.}
\label{fig:capio_impl}
\end{figure}

Device memory mapping is managed through CheriBSD's character device pager interface. 
It uses the region information provided by the kernel stub to create virtual memory objects, and in turn handle page faults for demand paging of the device memory regions. 
The memory mapping logic is sufficiently general for common device memory patterns such as MMIO registers, command queue descriptors, and device buffers.

\subsection{e1000e Kernel Stub}
The purpose of the kernel stub is to allocate DMA memory, handle the initial setup of the device, and define the shared memory regions that will be exposed to the \CAPIO{} interface. 
In the case of the e1000e this is handling the PCI probe, attach, and detach sequence, as well as allocating memory \callout{1} for the receive and transmit buffers and their corresponding descriptor rings. 
Critically, due to granular byte protections, instead of mapping the whole MMIO region \callout{2}, we can define individual slices for each MMIO register with their own unique permissions that are returned to the user \callout{4}. 
For example, the STATUS register is defined as readonly, but the CTRL register is not. 
In the case of receive and transmit descriptors \callout{3}, we can carve out the physical address portion of the descriptor struct  \callout{5}. 
This prevents arbitrary memory access with the device, but userspace can still modify the flags as needed in the transmit function, for example. 
To avoid going to the kernel for every receive and transmit, the descriptor rings are preprogrammed to a corresponding device buffer. 
Additionally, an ioctl call is exposed to allow for a privileged write to allowable regions. 
This could be updating the address in the transmit descriptor to point to a buffer the userspace process owns. 
In this case, the stub would verify that this is a valid address by checking the tag and extracting the physical address.

\subsection{Userspace Network Driver}

We developed a userspace driver for the Intel 82754L (e1000e) Gigabit Ethernet controller. This was built in 601 lines of C code. The driver attaches to the device through the \CAPIO{} ioctl interface and then makes mmap requests through the Slicer to map in the device memory. The life cycle of the device proceeds similarly to a typical e1000e driver. It implements LwIP's Network Interface (NETIF) API, where on transmit it copies the packet buffer to the next free transmit buffer, updates the tail descriptor, and then writes to the tail register. Receiving packets occurs on a separate thread that polls the ``done'' bit of the receive descriptor until is set and then loops through the available descriptors creating packet buffers for each and consuming them with NETIF's input function. Finally, it updates the tail to the last ready descriptor in the queue.

\subsection{LwIP}
\label{sec:impl-lwip}

To demonstrate that safe kernel bypass can support full-featured networking, we ported the LwIP \cite{lwip} network stack to the AArch64 architecture extended with Morello capability support. We used LwIP specifically for its broad support and ease of portability. Our focus is not on the performance of the network stack, but rather on demonstrating that \CAPIO{} can run real world applications, which require fully featured networking.

The primary challenge was retrofitting LwIP's memory model to adhere to CHERI's 128-bit capability width and strict pointer provenance rules. 
We had to ensure strict 128-bit alignment for all data structures containing pointers. Standard LwIP packet buffers and control blocks rely on packed structures that violate CHERI alignment requirements. We modified the LwIP configuration to enforce 128-bit alignment for all memory pools. 
The stack connects to our userspace driver through the NETIF API. To minimize latency, we copy LwIP's internal packet buffers directly to the device's buffers.

\begin{table}[t]
\centering
\caption{Code changes by component reflecting the overall porting effort for \CAPIO{} integration.}
\label{tab:code}

\begin{tabular}{l r}
\toprule
\textbf{Component} & \textbf{Lines of Code} \\
\midrule
CheriBSD & 50 \\
Slicer & 416 \\
Userspace Driver & 601 \\
\CAPIO{} Interface & 657 \\
Kernel Stub Driver & 1200 \\
Memcached & 120 \\
LwIP & 829 \\
\midrule
\textbf{Total} & \textbf{3,873} \\
\bottomrule
\end{tabular}
\end{table}

\subsection{Applications}

To demonstrate real-world viability of \CAPIO{}, we ported Memcached \cite{memcached}, a widely deployed high-performance key-value store to run on top of our LwIP network stack in userspace.
During this process, we encountered two challenges in porting standard POSIX applications to our system: getting them to build in pure-capability mode and operate correctly with LwIP.

The unique challenge posed by the CHERI architecture is the strict alignment requirement for capabilities.
Memcached's slab allocator, occasionally generated 8-byte aligned pointers for structures.
On a standard 64-bit system this is standard, but our implementation of CHERI on the ARM Morello has 128-bit pointers, and requires 16-byte aligned structures.
Loading a 128-bit capability from an 8-byte aligned address triggers a hardware fault.
This is a problem that is fundamental to CHERI, and required minimal changes to Memcached (see Table~\ref{tab:code}). 
The primary changes consisted of padding structures to satisfy alignment requirements.

 Memcached relies heavily on \texttt{libevent} to handle asynchronous I/O.
 Because our architecture bypasses the kernel network stack, standard event notification mechanisms such as \texttt{kqueue} or \texttt{epoll} cannot be used to monitor userspace sockets.
 To enable compatibility, we modified \texttt{libevent} to replace its \texttt{epoll}-based event handling with \texttt{select}.
 This modification is specific to our current implementation and not fundamental to \CAPIO{}, as other userspace network stacks, such as Fstack \cite{fstack}, provide their own \texttt{epoll}-compatible interfaces.

Out of the box, CheriBSD supports numerous applications, such as databases, web servers, etc.,  that have already been ported to CHERI \cite{p-cheribsd_ports}. For applications that have not been ported, porting effort can be reduced by the use of static analysis tools \cite{p-static-cheri} to avoid extensive testing at runtime.

\subsection{Userspace Programming of DMA}
The current implementation of \CAPIO{} relies on the kernel to manage the physical addresses written in DMA registers, as allowing the user application to write their own physical addresses as send/receive buffers would permit unbounded access to all memory.
This approach is not, however, required for DMA programming, it is just how our specific driver works.
To permit the userspace to program DMA addresses, an approach similar to SUD (Safe Userspace Drivers) could be taken, where the device is programmed to use a certain IOMMU region to restrict the device's physical memory access~\cite{tio-boyd}.
This allows the user to program addresses within that IOMMU region freely.
While we don't explore this technique in this work, we see this as an obvious area of future research for enabling devices which require runtime-reconfiguration of DMA (or writing DMA addresses to a command queue) such as NVMe.

\section{Evaluation} 
\label{sec:evaluation}
We now evaluate the effectiveness of the \CAPIO{} system we have implemented in \secref{sec:implementation}.
Our goal in this evaluation is to show that the systems enabled by \CAPIO{} can take an off-the-shelf commodity device and enable safe, performant, and clean userspace drivers to be built.
{\bf With this evaluation we seek to answer the following questions:}

\begin{itemize}
    \item Does the interface defined by \CAPIO{} effectively control access to protected device registers while making it easy to implement device drivers in userspace using those registers? (\secref{sec:eval-safe})
    \item Does the enforcement of fine-grained capability protections introduce significant overhead in the I/O data path?
    (\secref{sec:eval-latency})
    \item What is the raw latency improvement of a simple network card userspace driver built on \CAPIO{}? (\secref{sec:eval-latency-lwip})

    \item How does the \CAPIO{}-based kernel-bypass networking approach improve large networked applications and their throughput? (\secref{sec:eval-memcached})
        \item Is the interface exposed by \CAPIO{} extensible and general, and could it be used in general purpose device driver access?  (\secref{sec:eval-general})
\end{itemize}

\subsection{Experimental Setup}
\label{sec:setup}
Our evaluation was performed on an ARM Morello Development Board running four CHERI-enabled Neoverse N1 ARMv8.2-A processors clocked at 2.5GHz in a single NUMA domain.
The CPU features 64KB L1, 1MB Private L2, and 1MB shared L3 caches and 16GB of DDR4 registered DRAM.
We installed an Intel 82754L (e1000e) Gigabit Ethernet card in the first PCIe Gen3 x16 slot on the board and drove it with three configurations:
\begin{itemize}
    \item {\bf Kernel}: Socket-based interface through the CheriBSD kernel.
    \item {\bf \CAPIO{}-raw}: Raw UDP send/receive operations on the e1000e in userspace through \CAPIO{} capabilities.
    \item {\bf \CAPIO{}-LwIP}: Driving the e1000e in userspace with \CAPIO{} through the LWIP embedded network stack (\secref{sec:impl-lwip}).
\end{itemize}

We connected the Morello's e1000e network card \textit{directly} to a Fedora-based Dell Workstation running an identical network card driven through the kernel's network stack. The workstation is equipped with a 3.2GHz Intel i5 quad-core processor and 16 GB of RAM. The workstation can generate workloads that saturate the 1Gbps link. All latencies and bandwidths are measured as round-trip times from this Fedora workstation.

\subsection{\CAPIO{} Correctly Enforces Fine-grain Protections With Minimal Software Engineering}

 \label{sec:eval-safe}

By construction, CHERI capabilities are unforgeable \cite{CHERI, morello-validation, rigorous-engineering}. 
In a pure capability system, memory access without a capability is impossible. 
Furthermore, a process cannot access memory outside the bounds of a given capability. 
Therefore, if a memory region is mapped into their virtual memory space, a process can only read memory in that memory region if it possesses the appropriate capability. 
\CAPIO{} ensures that only the capabilities for the sliced and bounded memory regions are returned to the user, eliminating the risk of unprivileged access.

In \CAPIO{}, the kernel acts as the sole authority for capability derivation, ensuring that userspace drivers receive only the precise capabilities required for their operation. 
While the underlying physical page containing both safe and sensitive registers is mapped into the process's address space, the hardware prevents access to any byte not explicitly covered by a valid capability.

To validate that the \textit{Slicer} correctly enforces the policy defined in the \textit{Device Manifest}, we inspect the properties of the capabilities granted by it. 
We inspect three distinct registers: \texttt{CTRL} (Device Control), \texttt{STATUS} (Device Status), \texttt{IMS} (Interrupt Mask Set/Read) and \texttt{TDT} (Transmit Descriptor Tail).
An example manifest can be found in Table~\ref{tab:manifest}.
This manifest contains several registers which are marked as read+write, read-only, or kernel-only.
The IMS register is marked kernel only as it controls interrupt-related details which the user should not have access to.
We can verify this behavior by attempting to load the capability slices in userspace and inspecting their metadata using the CHERI API, as shown in the following implementation.

\begin{table}[t]
\centering
\caption{An Example Manifest for the e1000e}
\label{tab:manifest}
\begin{tabular}{l|l|c|l}
\toprule
\textbf{Register} & \textbf{Offset} & \textbf{Size} & \textbf{Permissions} \\
CTRL   & \texttt{0x0000} & 4 & RW \\
STATUS & \texttt{0x0008} & 4 & RO \\
IMS    & \texttt{0x0xD0} & 4 & Kernel \\
TDT    & \texttt{0x3818} & 4 & RW \\
\bottomrule
\end{tabular}
\end{table}

\begin{lstlisting}[language=C, 
    caption={Accessing the slices from the manifest in userspace is easy. They are directly mapped into capabilities.}, 
    label={lst:cap_inspection},
    basicstyle=\footnotesize\ttfamily,
    showstringspaces=false, 
    xleftmargin=0.5em,
    columns=fixed,
    breaklines=true,
    breakindent=1em,
    keywordstyle=\color{blue}\bfseries\small\ttfamily,
    stringstyle=\color{red!80!black}\small\ttfamily,
    emph={cheri_getbase, cheri_getlen, cheri_getperm, slices, __capability, slice_def_t},emphstyle=\color{purple}\bfseries\small\ttfamily]
/* Userspace Driver */
slices = map_mmio_slices()
enum Regs { REG_CTRL, REG_STATUS, ... };
print_capability("CTRL",   slices[REG_CTRL])
print_capability("STATUS", slices[REG_STATUS])
print_capability("TDT",    slices[REG_TDT])
\end{lstlisting}

The resulting output confirms the granularity of the protection mechanism. 
Notably, the \texttt{CTRL} and \texttt{STATUS} registers, located at offsets \texttt{0x00} and \texttt{0x08} respectively (Mapped in userspace a \texttt{0x40add000}, are distinct, non-overlapping capabilities).

\begin{lstlisting}[label={lst:cap_inspection_out},
    caption={The capabilities loaded from the slices table directly encode the manifest, and the hardware enforces this protection with unforgeable guarentees.}, 
    basicstyle=\footnotesize\ttfamily,
    showstringspaces=false, 
    xleftmargin=0.5em,
    columns=fixed,
    breaklines=true,
    breakindent=1em,
    keywordstyle=\color{blue}\bfseries\small\ttfamily,
    stringstyle=\color{red!80!black}\small\ttfamily,
    emph={cheri_getbase, cheri_getlen, cheri_getperm, __capability, slice_def_t},emphstyle=\color{purple}\bfseries\small\ttfamily]
CTRL:   0x40add000, len=4, Read+Write
STATUS: 0x40add008, len=4, Read Only 
TDT:    0x40ae0818, len=4, Read+Write
\end{lstlisting}

Despite residing on the same 4KB physical page, the hardware enforces strictly disparate policies for each register. 
The \texttt{STATUS} register is correctly identified as read-only and bounded to exactly 4 bytes, mirroring the manifest. 
Notably, the \texttt{IMS} register which is marked as kernel-only is \textit{not} present in the slices list, and thus it is impossible to access it.
This level of granularity confirms that \CAPIO{} successfully achieves sub-page safety. 
This protection guarantee can be further shown by attempting to offset from the CTRL register at offset 0 to the kernel-only interrupt mask register at offset \texttt{0xD0}:

\begin{lstlisting}[language=C, 
    caption={Attemping to write to the read-only STATUS register using an offset from the write-enabled control register is invalid.}, 
    label={lst:security_test},
    basicstyle=\footnotesize\ttfamily,
    xleftmargin=0.5em,
    columns=fixed,
    breaklines=true,
    breakindent=1em,
    commentstyle=\color{teal!60}\small\ttfamily,
    keywordstyle=\color{blue}\bfseries\small\ttfamily,
    stringstyle=\color{red!80!black}\small\ttfamily,
    emph={__capability},emphstyle=\color{purple}\bfseries\small\ttfamily]
u32 *ctrl = slices[REG_CTRL]; // Offset 0x0
u32 *ims = (u32*)((u8*)ctrl + 0xD0);
*ims = 42; // Invalid Capability (bounds)
// [SIGNAL CAUGHT] SIGPROT (Capability Fault)
\end{lstlisting}

The CHERI capability system provides strong guards against this kind of attack, and make it impossible to access memory outside of the range allowed in a given capability, even though the reside on the same page.

Despite these strict protections, the engineering effort required to develop a \CAPIO{} userspace driver is not significantly higher than that of a standard in-kernel driver. 
Because capabilities fundamentally behave as pointers, the core driver logic (reading status bits, writing commands, and updating ring buffers) remains unchanged. 
The primary distinction is the access mechanism: instead of calculating offsets from a single global base address, the driver indexes into the provided capability array. 
Consequently, the implementation complexity mirrors that of a traditional driver, allowing developers to focus on device logic rather than the underlying security enforcement.

\subsection{\CAPIO{} Enables Low-Latency I/O on Commodity Hardware}

\label{sec:eval-latency}

One of the main reasons to bring a device driver into userspace -- bypassing the kernel -- is to reduce end-to-end latencies.
Many systems built on top of kernel-bypassed network cards operate without a network stack, and often construct raw Ethernet frames in the buffers before sending them.
To that end, we evaluate the ``raw'' performance of our commodity e1000e network card using the \CAPIO{}-enabled userspace driver, comparing against the CheriBSD/FreeBSD network stack driven over the POSIX socket interface.

To measure the raw performance we ran a simple ``echo server'' on the ARM Morello using both \CAPIO{}-raw and kernel sockets.
The Dell workstation then sends UDP packets through the Linux network stack, varying two dimensions: packet size and inter-packet delay, measuring the time it takes for the Morello to echo that packet back.
\CAPIO{}-raw means the e1000e network card is driven by writing Ethernet+IP+UDP\footnote{We use UDP here as it has the lowest protocol overhead while still allowing the Dell workstation to receive the packets in userspace.} frames directly into the transmit buffers and by writing the descriptor index into the transmit queue.

Packets are then received by \textit{polling} on the tail descriptor ready bit on the e1000e card, and packets are read directly out of the transmit buffers.
This approach to using the network card should approach the lower bound in latencies achievable by the NIC, and this level of control is typical of latency-critical kernel bypass networking.

\begin{figure}[h]
    
    \includegraphics[width=1.0\columnwidth]{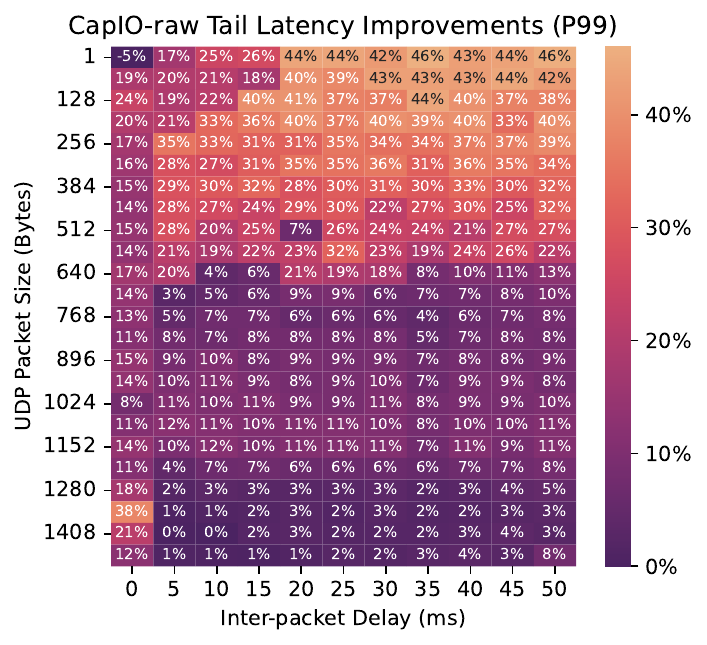}
    \caption{
    When a \CAPIO{}-enabled kernel-bypass network driver is driven directly with UDP packets, tail (here, the 99th percentile tail) round-trip latencies are reduces substantially - up to 46\% with packet sizes below 512 bytes. Higher numbers are better (lower latencies). These results are in-line with what is expected from kernel-bypass networking. The 46\% latency reduction is a reduction of $73\mu{}s$ ($235\mu{}s \rightarrow 162\mu{}s$).}
    \label{fig:capio-raw-rtt}
\end{figure}

The results of this are shown in Figure~\ref{fig:capio-raw-rtt}.  This figure, and others like it, constitute a heat map.   The vertical axis is the packet/write size (between 1 bytes and the MTU of 1472), the horizontal axis is the interval between packets/writes, and the heat dimension (color) is the improvement of the metric (99th percentile tail latency) of the \CAPIO{} version over the baseline version (higher percentages mean lower latencies).

The main takeaway here is that, as expected, bypassing the kernel improves latency considerably, by up to 46\% in the extremes, and the safety guarantees made by \CAPIO{} enable this latency gain on commodity hardware never designed for this kind of interface.
The latency gains seen by safe kernel-bypass are most apparent in situations where the fixed cost of the kernel dominate the overall time.
Those fixed kernel costs are, briefly: entering and exiting the kernel at least twice for the \verb|recvfrom| and \verb|sendto| system calls (roughly 450 cycles each), the kernel handling an interrupt from the NIC when packets arrive (and depart), and the cost of potentially scheduling and context switching back to the application using the network.
When there is no inter-packet delay, these costs dominate, and are easily combated with a driver whose operation can be contained entirely in userspace.
As soon as there is a delay between packets (5ms and above in Figure~\ref{fig:capio-raw-rtt}), this advantage is diminished as now the majority of the time spent handling a packet is simply waiting for it to arrive.

\subsection{Safe Userspace Networking is Practical}
\label{sec:eval-latency-lwip}

While the raw device access results in \secref{sec:eval-latency} demonstrate \CAPIO{}'s ability to eliminate kernel overhead, real-world applications rarely interact with network hardware at such a low level.
Most networked applications require a TCP/IP stack to handle protocol processing, connection management, and data segmentation.
The critical question is whether \CAPIO{}'s safety guarantees impose prohibitive overhead when integrated with a full network stack, and whether the performance benefits of kernel bypass survive the addition of userspace protocol processing.

To answer this question, we evaluated \CAPIO{} using LwIP, a lightweight IP stack originally designed for embedded systems.
LwIP prioritizes portability, low memory footprint, and implementation simplicity over raw performance.
Nothing about \CAPIO{} limits it to LwIP, and many fast userspace network stacks are available~\cite{fstack,dpdk,mtcp}.
LwIP contrasts sharply with our baseline, FreeBSD's network stack, which is known for its high performance~\cite{freebsd_performance} and serves as the foundation for systems like F-Stack~\cite{fstack}.
It has been tuned over decades for server workloads and includes sophisticated optimizations including zero-copy socket buffers.
By comparison, LwIP lacks many of the optimizations, and processes packets with significantly more memory copies and less efficient buffer management.

\begin{figure}[h]
    
    \includegraphics[width=1.0\columnwidth]{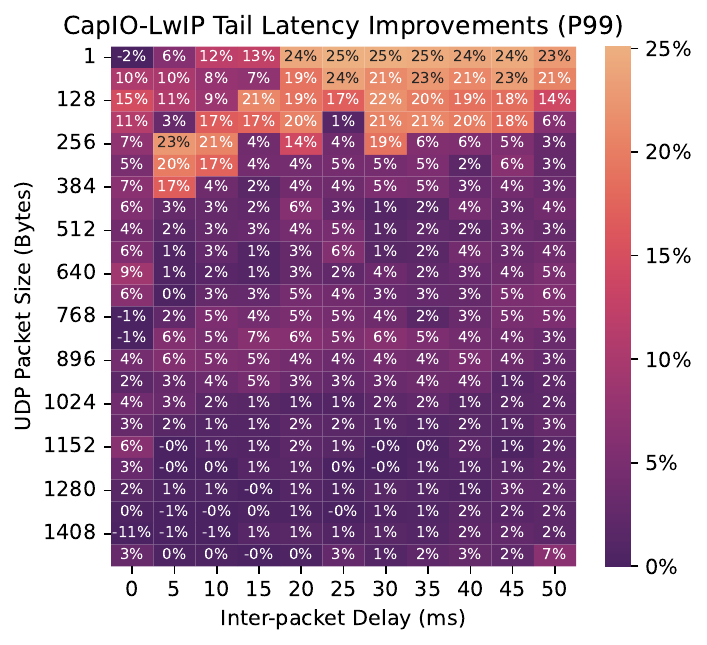}
    \caption{
    With a full network stack (LWIP), \CAPIO{}-based kernel bypass networking still
    outperforms the FreeBSD kernel's stack, but the margin is less. This is an improvement of $44\mu{}s$ over the kernel in the top right configuration in the plot, implying the LwIP network stack has roughly $30\mu{}s$ overhead in that configuration.
    }
    \label{fig:capio-lwip-rtt}
\end{figure}

We evaluate this LwIP implementation on top of a safe \CAPIO{}-enabled driver in an identical manner to \secref{sec:eval-latency}.
Figure~\ref{fig:capio-lwip-rtt} demonstrates that \CAPIO{}-enabled kernel-bypass with LwIP consistently reduces 99th percentile tail latencies compared to FreeBSD's kernel-mediated socket interface.
The improvements are most pronounced for small packets (16-300 bytes) at low inter-packet delays where we observe 15-25\% reduction in P99 latency.
This regime is particularly relevant for latency-sensitive applications like microservices, HPC, and real-time communication where small control messages dominate.

The magnitude of improvement is lower than the raw device access case (\secref{sec:eval-latency}) which is expected.
Having LwIP in the data path introduces additional overheads: packet copies between the network stack and application buffers, protocol header processing, checksum computation, and TCP state management.

As packet sizes increase beyond 300 bytes and inter-packet delays grow, the performance gap narrows.
This convergence occurs because larger packets amortize fixed per-packet costs (such as system call overhead) over more bytes, and longer delays mean that waiting for packet arrival dominates processing time.
In these scenarios, the absolute time spent in the kernel becomes less significant relative to the overall transaction time.

An important observation from these results is that \CAPIO{} decouples safety from performance.
The consistent improvements across packet sizes and delay configurations demonstrate that \CAPIO{}'s architectural approach to safe user-level access to byte-granular privileged memory does not adversely affect round-trip packet/write latencies.
We believe that integrating an alternative network stack such as the one found in F-stack would improve these results even further, but is outside the scope of this paper.

\subsection{Applications Using \CAPIO{}-based Kernel-Bypass Networking Benefit Similarly}
\label{sec:eval-memcached}
To demonstrate the real-world performance of \CAPIO{}, we ported Memcached \cite{memcached} to use the \CAPIO{}-based LwIP network stack from \secref{sec:eval-latency-lwip}.

Memcached is a widely deployed in-memory key-value store that serves as a critical component in web infrastructure, caching database query results, session state, and renderered page fragments.
It often handles millions of requests per second with strict latency requirements.
Its performance characteristics are well studied, and serves as a representative workload for latency sensitiveservices.

\begin{figure}[h]
    \includegraphics[width=0.9\columnwidth]{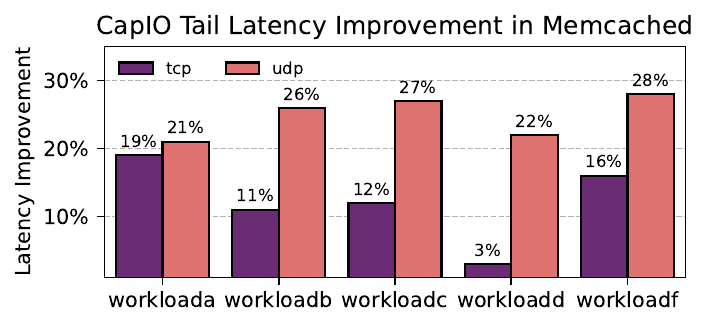}
    \caption{
        End-to-end latencies in Memcached driven by the YCSB workload generator have improvements across the board in TCP and UDP.
    }
    \label{fig:memcached-ycsb}
\end{figure}

We evaluated this modified version of memcached using the Yahoo Cloud Serving Benchmark suite (YCSB) \cite{COOPER-YCSB-2010}, an open-source designed to evaluate the performance of key-value and cloud data serving systems under realistic workloads.
It provides a standardized set of workload patterns ranging from read-heavy (workloadb), write-heavy (workloada) to read-modify-write scenarios (workloadf).
We used YCSB to generate workloads and evaluated these against memcached using both TCP and UDP protocols\footnote{Note that the UDP protocol for memcached is not fully supported by the \textit{libmemcached} library, so some light engineering was needed to get it working.}.
The latency improvements are shown in Figure~\ref{fig:memcached-ycsb} for these workloads.
These results demonstrate that \CAPIO{}'s architectural benefits of sub-page device isolation through hardware capabilities and translate into measurable, practical improvements for real-world networked applications.

\subsection{The \CAPIO{} Interface is Generalizable}
\label{sec:eval-general}

Although we evaluated \CAPIO{} using a network interface card, \CAPIO{} is not specifically tailored to them. The enforcement mechanisms in \CAPIO{}'s design, such as the Slicer, are applicable to many devices. The fundamental problem addressed by \CAPIO{}, namely the protection granularity problem, is due to the close proximity of sensitive control plane registers with data plane registers, a proximity that appears in the interface of many MMIO devices.  This occurs regularly in the design of devices because hardware designers often prioritize performance and simplicity over strict isolation.  It also occurs when devices must be mappable both as MMIO and as port-mapped I/O and the I/O port address space is very small (e.g., for x64 systems).

Most device interfaces are similar to the device interface of the e1000e card we used in our evaluation. These device interfaces are typically structured around a command queue with descriptors. The device driver prepares descriptors, which are data structures that describe operations that are to be performed, such as packet transmissions or disk reads, and places them into a command queue that the device continuously monitors. Each descriptor contains metadata, such as buffer locations, data lengths, and control flags, which allow the device to process the commands without additional host intervention. 

This is popular in high throughput devices because it enables the device to processes multiple commands in parallel.  It also makes adapting drivers for such devices to the \CAPIO{} model particularly easy.

\section{Conclusion and Future Work} 
\label{sec:conclusion}

We introduced \CAPIO{}, the first architecture to leverage hardware capabilities to enforce fine-grained access control on MMIO. This is achieved by using capabilities to create precise, sub-page ``slices'' of MMIO regions. \CAPIO{} enables the kernel to delegate latency-critical hardware access to userspace applications, and preventing access to privileged registers that are on the same memory page. 

We implemented \CAPIO{} on the ARM Morello platform with a commodity network card and showed that we can achieve this fine-grained memory access with minimal performance overhead. We evaluated \CAPIO{} on microbenchmarks and a real world application, Memcached. Microbenchmarked UDP shows a 46\% decrease in latency and application TCP latency show a decrease by up to 19\% decrease in latency. Future work includes exploring the use of \CAPIO{} on NVMe storage devices and other privileged memory such as exposing raw access to kernel datastructures to expand the functionality of interfaces like VDSO.

\bibliographystyle{IEEEtran}
\bibliography{references}

@inproceedings{COOPER-YCSB-2010,
	author = {Cooper, Brian F. and Silberstein, Adam and Tam, Erwin and Ramakrishnan, Raghu and Sears, Russell},
	title = {Benchmarking Cloud Serving Systems with YCSB},
	year = {2010},
	isbn = {9781450300360},
	publisher = {Association for Computing Machinery},
	address = {New York, NY, USA},
	url = {https://doi.org/10.1145/1807128.1807152},
	doi = {10.1145/1807128.1807152},
	booktitle = {Proceedings of the 1st ACM Symposium on Cloud Computing},
	pages = {143–154},
	numpages = {12},
	location = {Indianapolis, Indiana, USA},
	series = {SoCC '10}
}

@ARTICLE{morello,
  author={Grisenthwaite, Richard and Barnes, Graeme and Watson, Robert N. M. and Moore, Simon W. and Sewell, Peter and Woodruff, Jonathan},
  journal={IEEE Micro}, 
  title={The Arm Morello Evaluation Platform—Validating CHERI-Based Security in a High-Performance System}, 
  year={2023},
  volume={43},
  number={3},
  pages={50-57},
  doi={10.1109/MM.2023.3264676}}

@INPROCEEDINGS{CHERI,
  author={Watson, Robert N.M. and Woodruff, Jonathan and Neumann, Peter G. and Moore, Simon W. and Anderson, Jonathan and Chisnall, David and Dave, Nirav and Davis, Brooks and Gudka, Khilan and Laurie, Ben and Murdoch, Steven J. and Norton, Robert and Roe, Michael and Son, Stacey and Vadera, Munraj},
  booktitle={2015 IEEE Symposium on Security and Privacy}, 
  title={CHERI: A Hybrid Capability-System Architecture for Scalable Software Compartmentalization}, 
  year={2015},
  volume={},
  number={},
  pages={20-37},
  doi={10.1109/SP.2015.9}}

@unpublished{nickgordon,
           title = {Secure I/O on trusted platforms with lightweight kernels.},
            year = {2025},
           month = {January},
          author = {Nicholas Gordon},
             url = {https://d-scholarship.pitt.edu/46961/}
}

@inproceedings{cheribsd,
  author = {Sean M. Seefried and Peter G. Neumann and Robert N. M. Watson and Jonathan Anderson and Alex Richardson and Mark D. Richardson and David Chisnall and Michael Roe and Andrew W. Appel and Robert N. M. Watson},
  title = {CheriBSD: A Research Fork of FreeBSD},
  booktitle = {AsiaBSDCon 2016},
  year = {2016},
  url = {https://papers.freebsd.org/2016/asiabsdcon/brooks-cheri/},
  note = {Conference paper discussing CHERI-enabled FreeBSD extensions and challenges}
}

@ARTICLE{tio-liang,
  author={Liang, Hongliang and Li, Mingyu and Chen, Yixiu and Jiang, Lin and Xie, Zhuosi and Yang, Tianqi},
  journal={IEEE Transactions on Information Forensics and Security}, 
  title={Establishing Trusted I/O Paths for SGX Client Systems With Aurora}, 
  year={2020},
  volume={15},
  number={},
  pages={1589-1600},
  doi={10.1109/TIFS.2019.2945621}}

@techreport{tio-amd,
  title        = {AMD SEV-TIO: Trusted I/O for Secure Encrypted Virtualization},
  institution  = {Advanced Micro Devices, Inc.},
  year         = {2023},
  note         = {Whitepaper},
  url          = {https://www.amd.com/content/dam/amd/en/documents/developer/sev-tio-whitepaper.pdf}
}

@inproceedings{tio-sgx,
  title={Sgxio: Generic trusted i/o path for intel sgx},
  author={Weiser, Samuel and Werner, Mario},
  booktitle={Proceedings of the seventh ACM on conference on data and application security and privacy},
  pages={261--268},
  year={2017}
}

@inproceedings{tio-guo,
  title={Minimum viable device drivers for arm trustzone},
  author={Guo, Liwei and Lin, Felix Xiaozhu},
  booktitle={Proceedings of the Seventeenth European Conference on Computer Systems},
  pages={300--316},
  year={2022}
}

@inproceedings{tio-boyd,
  title={Tolerating malicious device drivers in Linux},
  author={Boyd-Wickizer, Silas and Zeldovich, Nickolai},
  booktitle={2010 USENIX Annual Technical Conference (USENIX ATC 10)},
  year={2010}
}

@misc{dpdk,
  author = {{DPDK Project}},
  title = {DPDK – The open source data plane development kit},
  howpublished = {\url{https://www.dpdk.org}},
  year = {2025},
  note = {Accessed: 2025-12-06}
}

@INPROCEEDINGS{spdk,
  author={Yang, Ziye and Harris, James R. and Walker, Benjamin and Verkamp, Daniel and Liu, Changpeng and Chang, Cunyin and Cao, Gang and Stern, Jonathan and Verma, Vishal and Paul, Luse E.},
  booktitle={2017 IEEE International Conference on Cloud Computing Technology and Science (CloudCom)}, 
  title={SPDK: A Development Kit to Build High Performance Storage Applications}, 
  year={2017},
  volume={},
  number={},
  pages={154-161},
  doi={10.1109/CloudCom.2017.14}}

@inproceedings{netmap,
  title={netmap: a novel framework for fast packet I/O},
  author={Rizzo, Luigi},
  booktitle={21st USENIX Security Symposium (USENIX Security 12)},
  pages={101--112},
  year={2012}
}

@INPROCEEDINGS{cheri-ferraro,
  author={Ferraro, Donato and Bastoni, Andrea and Zuepke, Alexander and Marongiu, Andrea},
  booktitle={2025 Design, Automation \& Test in Europe Conference (DATE)}, 
  title={Enabling Security on the Edge: A CHERI Compartmentalized Network Stack}, 
  year={2025},
  volume={},
  number={},
  pages={1-7},
  doi={10.23919/DATE64628.2025.10992964}}

@misc{fstack,
  title        = {F-Stack: Userspace Network Development Kit},
  author       = {{Tencent}},
  howpublished = {\url{https://github.com/F-Stack/f-stack}},
  note         = {Accessed: 2025-12-08},
}

@inproceedings{bypassd,
  title={BypassD: Enabling fast userspace access to shared SSDs},
  author={Yadalam, Sujay and Alverti, Chloe and Karakostas, Vasileios and Gandhi, Jayneel and Swift, Michael},
  booktitle={Proceedings of the 29th ACM International Conference on Architectural Support for Programming Languages and Operating Systems, Volume 1},
  pages={35--51},
  year={2024}
}

@inproceedings {ipc_stuart,
author = {D. Stuart Ritchie and Gerald W. Neufeld},
title = {User Level {IPC} and Device Management in the Raven Kernel},
booktitle = {USENIX Microkernels and Other Architectures Symposium (USENIX Microkernels and Other Architectures Symposium)},
year = {1993},
address = {San Diego, CA},
url = {https://www.usenix.org/conference/usenix-microkernels-and-other-architectures-symposium/user-level-ipc-and-device},
publisher = {USENIX Association},
month = sep
}

@article{exokernel-engler,
author = {Engler, D. R. and Kaashoek, M. F. and O'Toole, J.},
title = {Exokernel: an operating system architecture for application-level resource management},
year = {1995},
issue_date = {Dec. 3, 1995},
publisher = {Association for Computing Machinery},
address = {New York, NY, USA},
volume = {29},
number = {5},
issn = {0163-5980},
url = {https://doi.org/10.1145/224057.224076},
doi = {10.1145/224057.224076},
journal = {SIGOPS Oper. Syst. Rev.},
month = dec,
pages = {251–266},
numpages = {16}
}

@article{high-perf-sriov,
  title={High performance network virtualization with SR-IOV},
  author={Dong, Yaozu and Yang, Xiaowei and Li, Jianhui and Liao, Guangdeng and Tian, Kun and Guan, Haibing},
  journal={Journal of Parallel and Distributed Computing},
  volume={72},
  number={11},
  pages={1471--1480},
  year={2012},
  publisher={Elsevier}
}

@inproceedings{mtcp,
author = {Jeong, Eun Young and Woo, Shinae and Jamshed, Muhammad and Jeong, Haewon and Ihm, Sunghwan and Han, Dongsu and Park, KyoungSoo},
title = {mTCP: a highly scalable user-level TCP stack for multicore systems},
year = {2014},
isbn = {9781931971096},
publisher = {USENIX Association},
address = {USA},
booktitle = {Proceedings of the 11th USENIX Conference on Networked Systems Design and Implementation},
pages = {489–502},
numpages = {14},
location = {Seattle, WA},
series = {NSDI'14}
}

@article{lwip,
  title={Design and Implementation of the lwIP TCP/IP Stack},
  author={Dunkels, Adam},
  journal={Swedish Institute of Computer Science},
  volume={2},
  number={77},
  year={2001}
}

@article{freebsd_performance,
  title={Performance characterization of the FreeBSD network stack},
  author={Kim, Hyong-youb and Rixner, Scott},
  journal={Computer Science Department, Rice University},
  year={2005}
}

@misc{memcached,
  author = {Brad Fitzpatrick},
  title = {Memcached: A distributed memory object caching system},
  year = {2003},
  howpublished = {\url{https://memcached.org/}},
  note = {Originally developed at Danga Interactive}
}

@inproceedings{kernel-bypass-is-not-optional,
author = {Jasny, Matthias and El-Hindi, Muhammad and Ziegler, Tobias and Binnig, Carsten},
title = {A Wake-Up Call for Kernel-Bypass on Modern Hardware},
year = {2025},
isbn = {9798400719400},
publisher = {Association for Computing Machinery},
address = {New York, NY, USA},
url = {https://doi.org/10.1145/3736227.3736235},
doi = {10.1145/3736227.3736235},
booktitle = {Proceedings of the 21st International Workshop on Data Management on New Hardware},
articleno = {14},
numpages = {5},
location = {
},
series = {DaMoN '25}
}

@manual{onload,
  title = {ef\_vi User Guide},
  author = {Xilinx, Inc.},
  organization = {Xilinx, Inc.},
  year = {2019},
  month = sep,
  number = {SF-114063-CD},
  edition = {Issue 10},
  url = {https://www.amd.com/content/dam/amd/en/support/downloads/solarflare/onload/enterprise-onload/SF-114063-CD-10_ef_vi_User_Guide.pdf},
  note = {Last Revised: September 2019; Accessed: 2025-12-10},
  howpublished = {\url{https://www.amd.com/content/dam/amd/en/support/downloads/solarflare/onload/enterprise-onload/SF-114063-CD-10_ef_vi_User_Guide.pdf}}
}

@misc{g-nvme2019spec,
  title = {NVMe Base Specification},
  author = {NVM Express, Inc.},
  year = {2019},
  edition = {1.4},
  note = {Submission and completion queues use descriptor rings for command processing, analogous to NIC TX/RX rings},
  url = {https://nvmexpress.org/specifications/}
}

@manual{g-vfio2023,
  title = {VFIO - ``Virtual Function I/O''},
  organization = {The Linux Kernel Documentation},
  year = {2023},
  note = {PCIe devices (NICs, storage, GPUs) expose uniform doorbells and descriptor rings for user-space DMA via mediated device model},
  url = {https://docs.kernel.org/driver-api/vfio.html}
}

@misc{g-spdk2024nvme,
  title = {SPDK NVMe Driver},
  organization = {Intel SPDK Project},
  author = {Intel},
  year = {2024},
  note = {NVMe poll-mode driver uses submission/completion queue pairs with descriptor rings, extending DPDK NIC patterns to storage},
  url = {https://spdk.io/doc/nvme.html}
}

@inproceedings{p-static-cheri,
author = {Dudina, Irina and Stark, Ian},
title = {Static Analysis for Transitioning to CHERI C/C++},
year = {2024},
isbn = {9798400706219},
publisher = {Association for Computing Machinery},
address = {New York, NY, USA},
url = {https://doi.org/10.1145/3652588.3663323},
doi = {10.1145/3652588.3663323},
booktitle = {Proceedings of the 13th ACM SIGPLAN International Workshop on the State Of the Art in Program Analysis},
pages = {52–59},
numpages = {8},
location = {Copenhagen, Denmark},
series = {SOAP 2024}
}

@misc{p-cheribsd_ports,
  author = {CTSRD-CHERI and FreeBSD Ports Team},
  title = {cheribsd-ports: FreeBSD ports tree adapted for CheriBSD},
  howpublished = {\url{https://github.com/CTSRD-CHERI/cheribsd-ports}},
  year = {2025},
  note = {Accessed: 2025-12-10; Latest commit: 64bcbb0 by kwitaszczyk on Apr 17, 2025},
  organization = {Computer Laboratory, University of Cambridge}
}

@ARTICLE{morello-validation,
  author={Grisenthwaite, Richard and Barnes, Graeme and Watson, Robert N. M. and Moore, Simon W. and Sewell, Peter and Woodruff, Jonathan},
  journal={IEEE Micro}, 
  title={The Arm Morello Evaluation Platform—Validating CHERI-Based Security in a High-Performance System}, 
  year={2023},
  volume={43},
  number={3},
  pages={50-57},
  doi={10.1109/MM.2023.3264676}}

@inproceedings{rigorous-engineering,
  title={Rigorous engineering for hardware security: Formal modelling and proof in the CHERI design and implementation process},
  author={Nienhuis, Kyndylan and Joannou, Alexandre and Bauereiss, Thomas and Fox, Anthony and Roe, Michael and Campbell, Brian and Naylor, Matthew and Norton, Robert M and Moore, Simon W and Neumann, Peter G and others},
  booktitle={2020 IEEE Symposium on Security and Privacy (SP)},
  pages={1003--1020},
  year={2020},
  organization={IEEE}
}

\end{document}